\documentclass[aps,prl,reprint,superscriptaddress]{revtex4-2}
\usepackage{amsmath,graphicx,color,epsfig}
\usepackage{graphicx,color,epsfig}
\usepackage{amssymb,amsmath,enumerate}

\begin{document}
\title{Chimera states and information transfer in interacting populations of map-based neurons}

\author{V. J. M\'arquez-Rodr\'iguez}
\affiliation{Laboratorio de Neurociencia Social y Neuromodulaci\'on, Centro de Investigaci\'on en Complejidad Social (neuroCICS), Facultad de Gobierno, Universidad del Desarrollo, Santiago, Chile}
\affiliation{Facultad de Ciencias, SUMA-CeSiMo, Universidad de Los Andes, M\'erida, Venezuela}
\author{K. Tucci}
\affiliation{Facultad de Ciencias, SUMA-CeSiMo, Universidad de Los Andes, M\'erida, Venezuela}
\author{M. G. Cosenza}
\affiliation{Grupo Interdisciplinario de Sistemas Complejos, Escuela de Ciencias F\'isicas y Nanotecnolog\'ia, Universidad Yachay Tech, Urcuqu\'i, Ecuador.}
\date{\sf Neural Computing  \& Applications (2024). https://doi.org/10.1007/s00521-024-10050-3}

\begin{abstract}
We investigate the synchronization behavior and the emergence of chimera states in a system of two interacting populations of maps possessing chaotic neural-like dynamics.
 We characterize four collective states  on the space of coupling parameters of the system: 
complete synchronization, generalized synchronization, chimera states, and incoherence.
We quantify the information exchange between the two neuron populations  in chimera states.
We have found a well-defined direction of the flow of information in chimera states, from the desynchronized  population to the synchronized one. The incoherent population functions as a driver of the coherent neuron population in a chimera state. This feature
is independent of the population sizes or population partitions. Our results yield insight into the communication mechanisms arising in brain processes such as unihemispheric sleep and epileptic seizures that have been associated to chimera states.
\end{abstract}
 \maketitle

\section{Introduction}

The study of emergent collective states in systems composed of 
interacting populations of dynamical units
is a relevant topic in complexity science. These systems have been investigated in many different areas, such as coupled networks of oscillators \cite{montbrio2004synchronization, abrams2008solvable}, cohabitation of two biological species \cite{goel1971volterra, gomatam36new}, 
competition of two languages \cite{patriarca2004modeling}, and neural networks \cite{stefanescu2008low,maslennikov2014modular}. 
In the context of neural systems, the study of interacting populations of dynamical units can be relevant to understand the functional dynamics of the brain  \cite{cardanobile2011emergent,magyar2015two}. Two-population models 
have been used in several studies, such as
synchronization between oscillations emerging from separated cortical neuronal assemblies \cite{bibbig2002long}, neuronal information processing \cite{battaglia2012dynamic}, and phase-coherence transitions between delay-coupled neuronal populations \cite{barardi2014phase}.

In many systems, collective behavior can be  characterized as synchronization states arising from the interactions within and between populations. 
Recently, there has been great interest in the study of chimera states in dynamical networks \cite{kuramoto2002coexistence,abrams2004chimera,panaggio2015chimera,zakharova2020chimera}. A chimera state consists of the coexistence of subsets of elements with synchronous and asynchronous dynamics in a spatiotemporal system. In a system of two interacting populations, a chimera state
is manifested as one population displaying a synchronized behavior while the other remains desynchronized. Chimera states have been found in systems consisting of two interacting populations of oscillators \cite{abrams2008solvable,laing2010chimeras,omelchenko2011loss,martens2016chimera,premalatha2017chimeralike}, and in cross-cultural interactions of two social groups \cite{gonzalez2014localized}. They have been experimentally observed in two coupled populations of mechanical oscillators \cite{martens2013chimera} and electrochemical oscillators  \cite{tinsley2012chimera}.

Chimera states have been associated to brain processes such as unihemispheric sleep in various animal species including birds, aquatic mammals, and reptiles \cite{rattenborg2000behavioral,lesku2009phylogeny}, as well as human electroencephalographic patterns in epileptic seizures \cite{lainscsek2019cortical,Andrzejak2016}. 
Chimera states have been found in a two-layer network brain  model based on data from cerebral cortex \cite{kang2019two} and in two-layer neuronal network with unidirectional inter-layer links \cite{li2019synchronization}. Chimeras in neural systems have been mostly studied in models of coupled differential equations, such as the Hodgkin-Huxley model \cite{glaze2016chimera}, the Hindmarsh-Rose model \cite{hizanidis2016chimera,bera2016chimera,glaze2021neural}, and the Fitzhugh-Nagumo model \cite{omelchenko2015robustness,essaki2015chimeric}. More recently, chimera states have been reported in networks with neural-type local dynamics described by the Rulkov time-discrete map \cite{rybalova2019spiral,mehrabbeik2021synchronization}.

In many situations it is important not only to characterize chimeras or other collective states, but also to understand
the causal relationships between the constituent parts of a system that lead to such behaviors. 
In particular, information transfer measures have been proven useful
 to quantify drive-response causal relationships between subsystems
and functional structures
in diverse complex systems \cite{TE}.
For example,  the emergence of nontrivial collective behavior in chaotic dynamical networks has been associated to the flow of information from global to local scales \cite{cisneros2002information}.
Transfer entropy methods have been widely employed in neuroscience to evaluate  interdependence between electroencephalographic data sets \cite{wibral2014transfer}.  Such measures allow, for instance, to evaluate coupling directions  \cite{li2020measuring} and connectivity \cite{ursino2020transfer} between different regions of the brain.

In this article, we investigate the emergence of  synchronization and chimera states in a system of two interacting populations of chaotic maps possessing neural-like dynamics. We characterize various synchronization states that arise in the system: complete synchronization, generalized synchronization, chimera states, and incoherence. Specifically, we address the question: who is the driver in a chimera state in neuron dynamical networks? We quantify the
information flow between the synchronized and desynchronized neuron populations in a chimera state by employing the information transfer
measure of Schreiber \cite{Schreiber}
in order to gain insight into the 
communication mechanisms associated to 
pathologies in the brain. We compare the information transfer between the mean fields of the two populations in chimera states.
Our approach is simpler than methods based on delayed mutual information and Poincar\'e sections used to detect the flow of information in chimera states of  phase oscillator networks \cite{Deschle2019}.
 
In Sec.~2, we introduce a model of two populations interacting through their mean fields and define the order parameters to characterize synchronization states. Section~3 describes the spatiotemporal patterns associated to the different collective synchronization states arising in the system. The collective states are characterized on the phase space given by the coupling parameters of the system. Section~4 contains the calculation of the information transfer between the two neuron populations in a chimera state. We find a definitive direction of the flow of information from the desynchronized population to the synchronized one. Conclusions are presented in Sec.~5.

\section{Model for two interacting populations of map-based neurons}
Coupled map lattices or coupled map networks are spatiotemporal dynamical systems
where space and time are discrete, but the state variables are continuous. They consist of a set of maps or iterative functions considered as nodes interacting on a lattice or network \cite{kaneko1984period,waller1984spatial}. Coupled map networks have provided useful models for the study of diverse processes in spatially extended systems, with the advantage of being computationally efficient \cite{kaneko1993theory}.  The discrete-space character of coupled map systems makes them  appropriate for the investigation of dynamics on nonuniform networks \cite{cosenza1992coupled}.  

Our system is composed of $N$ maps possessing neuron dynamics, distributed into two populations denominated as $\alpha$ and $\beta$, with sizes $N^\alpha$ and $N^\beta$, respectively, such that $N=N^\alpha+N^\beta$. In order to model local excitable dynamics, we consider two-dimensional map-based neurons where the two variables may represent the membrane potential and outward ionic currents, respectively. We use the notation $[k]$  to indicate ‘‘or k’’. Then, the state of element $i[j] \in \alpha [\beta]$ at  discrete time $t$ is given by two variables $x_t^\alpha(i),y_t^\alpha(i)$ $[x^\beta_t(j),y_t^\beta(j)]$, where $i=1,2,\ldots, N^\alpha$; $j=1,2,\ldots, N^\beta$. We assume that each element within a population interacts with the mean field of that population and with the mean field of the other population. 
Mean-field coupling has been  used in neural mass models \cite{stefanescu2008low,naze2015computational,breakspear2007neuronal}. Then, we define the  dynamics of the two interacting populations of map-based neurons by the following coupled map equations,
\begin{eqnarray}
\label{ga1}
x^{\alpha}_{t+1}(i) &=& (1 - \mu) f( x^{\alpha}_t(i) , y^{\alpha}_t(i)) +   \mu \bar X_t^\alpha + \epsilon \bar X_t^\beta, \\
y^{\alpha}_{t+1}(i) &=& g( x^{\alpha}_t(i) , y_t^{\alpha}(i) ); \\
\label{ga2}
\label{gb1}
x^{\beta}_{t+1}(j) &=& (1 - \mu) f( x^{\beta}_t(j) , y_t^{\beta}(j) ) + \mu \bar X_t^\beta + \epsilon \bar X_t^\alpha,  \\
y^{\beta}_{t+1}(j) &=&  g ( x^{\beta}_t(j) , y_t^{\beta}(j) ),
\label{gb2}
\end{eqnarray}
where the functions $f(x,y)$ and $g(x,y)$ describe the local dynamics, and parameters $\mu$ and $\epsilon$ characterize the strength of the intra-population and inter-population coupling, respectively. The mean fields of populations $\alpha$ and $\beta$ at time $t$ are defined, respectively, as
\begin{eqnarray}
\label{CMxa}
\bar X_t^{\alpha} &=& \dfrac{1}{N^{\alpha}} \sum_{i=1}^{N^\alpha} x^{\alpha}_t(i), \\
\bar X_t^{\beta} &=& \dfrac{1}{N^{\beta}} \sum_{j=1}^{N^\beta} x^{\beta}_t(j).
\label{CMxb}
\end{eqnarray}

\begin{figure}[h]
\centerline{\includegraphics[scale=0.6]{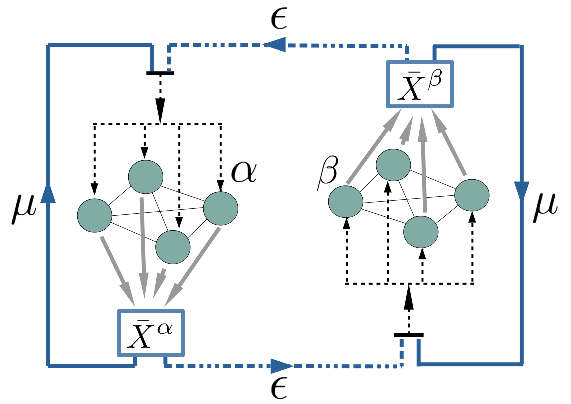}}
\caption{Scheme of two populations $\alpha$ and $\beta$ reciprocally interacting through their mean fields. Any element in population $\alpha[\beta]$ interacts with: (i) all other elements in $\alpha[\beta]$ through the mean field $\bar X^\alpha$ $[\bar X^\beta]$ with coupling parameter $\mu$, and (ii) the mean field $\bar X^\beta$ $[\bar X^\alpha]$ of population $\beta[\alpha]$ with strength $\epsilon$.}
\label{scheme}
\end{figure}

Figure~(\ref{scheme}) illustrates the system of two interacting populations of dynamical elements Eqs.~(\ref{ga1})-(\ref{gb2}).

Synchronization states for each population can be characterized by the following asymptotic time averages, after discarding a number of transients $\tau$,

\begin{eqnarray}
\langle \sigma^{\alpha} \rangle  &=& \frac{1}{T- \tau} \sum_{t=\tau}^T \sigma_t^{\alpha}, \\
\langle \sigma^{\beta} \rangle  &=& \frac{1}{T- \tau} \sum_{t=\tau}^T \sigma_t^{\beta}, \\
\langle \delta \rangle &=& \frac{1}{T- \tau} \sum_{t=\tau}^T |\bar X_t^{\alpha} - \bar X_t^{\beta}|,
\end{eqnarray}
where the instantaneous standard deviations of the distribution of state variables are defined by
\begin{eqnarray}
\sigma_t^{\alpha} &=& \left[ \frac{1}{N^\alpha} \sum_{i=1}^{N^\alpha} (x^{\alpha}_t(i) - \bar X_t^{\alpha})^2 \right]^{1/2}, \\
\sigma_t^{\beta} &=& \left[ \frac{1}{N^\beta} \sum_{j=1}^{N^\beta} (x^{\beta}_t(j) - \bar X_t^{\beta})^2 \right]^{1/2}.
\end{eqnarray}
A collective state of synchronization within population $\alpha$ $[\beta]$ for the system represented by equations (\ref{ga1})-(\ref{ga2}) and (\ref{gb1})-(\ref{gb2}) takes place when $x^{\alpha[\beta]}_t(i) = x^{\alpha[\beta]}_t(k)=\bar X_t^{\alpha[\beta]}$ and $y^{\alpha[\beta]}_t(i) = y^{\alpha[\beta]}_t(k)$, $\forall i, k \in \alpha [\beta]$, sustained in time. 
Thus, a complete state of synchronization in population $\alpha$ is characterized by the condition $\langle \sigma^{\alpha} \rangle = 0$, and similarly population $\beta$ is synchronized when $\langle \sigma^{\beta} \rangle = 0$. In numerical simulations, we set the criterion $\langle \sigma^{\alpha[\beta]} \rangle < 10^{-7}$ for synchronization of a population.

Different collective states of synchronization can be defined in the system Eqs. (\ref{ga1})-(\ref{gb2}):
\begin{enumerate}
\item {\it Complete synchronization} (CS): when all elements within each population are synchronized with each other and with the elements of the other population. That is, $\langle \sigma^{\alpha} \rangle = 0$, $\langle \sigma^{\beta} \rangle = 0$, and $\langle \delta \rangle = 0$.
\item {\it Generalized Synchronization} (GS): when the elements within each population are synchronized but not with the elements of the other population \cite{AC}. That is, $\langle \sigma^{\alpha} \rangle = 0$ and $\langle \sigma^{\beta} \rangle = 0$, but  $\langle \delta \rangle \neq 0$.
\item {\it Chimera state} (Q): when the elements in one population are synchronized but the elements in the other population are not. That is, $\langle \sigma^{\alpha} \rangle = 0$, $\langle \sigma^{\beta} \rangle \neq 0$, or viceversa, and $\langle \delta \rangle \neq 0$.
\item {\it Desynchronization} (D): none of the populations exhibits synchronization. This means, $\langle \sigma^{\alpha} \rangle \neq 0$ and $\langle \sigma^{\beta} \rangle \neq 0$.
\end{enumerate}

We consider the two-dimensional map proposed by Rulkov \cite{rulkov2002modeling} as local neural dynamics, defined as
\begin{eqnarray}
\label{xrul}
x_{t+1} &=& h(x_t, y_t), \\
y_{t+1} &=& y_t - \upsilon(x_t+1) + \upsilon \gamma ,
\label{yrul}
\end{eqnarray}
where $x_t$ and $y_t$ are the fast and slow variables. As for the two previous local maps, although these variables have not specific biological meaning, $x_t$ can be interpreted as the transmembrane potential of a neuron, while $y_t$ is a recovery or adaptation variable, where slow time evolution is given by small values of the parameter $\upsilon$ $(0<\upsilon \ll 1)$.

\begin{figure}[h]
\centerline{\includegraphics[width=7cm]{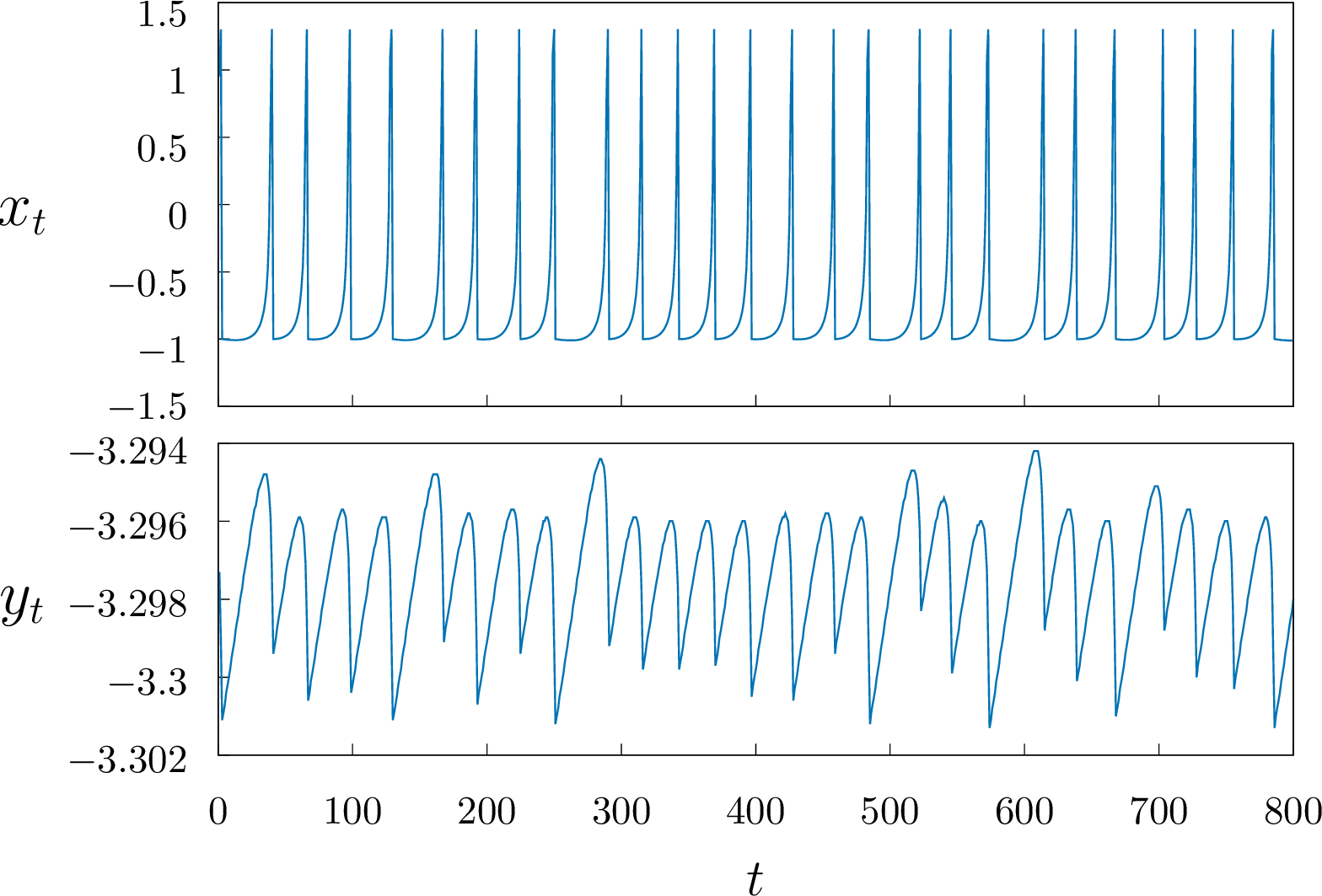}}
\caption{Time series of a single Rulkov map, Eqs.~(\ref{xrul})-(\ref{yrul}), with parameters $\upsilon=0.001$, $\varrho=4.6$, $\gamma=0.225$.}
\label{rulkovseries}
\end{figure}

To generate spiking and silent regimes, Eq.~(\ref{xrul}) employs a piecewise function $h(x,y)$ of the form 
\begin{align}
h(x,y) = \left\{ \begin{array}{lcc}
             \varrho(1-x)^{-1} + y  &   \textrm{if}  & x \leq 0, \\
              \varrho + y &  \textrm{if} & 0 < x < \varrho + y, \\
	      -1 &  \textrm{if} &  x \geq \varrho + y.            \end{array}
   \right.
\label{heavisider}
\end{align}
where $\gamma$ and $\varrho$ are control parameters that allow to select different regimes of temporal behavior of the model. 
Figure~\ref{rulkovseries} shows the time evolution of the Rulkov map variables for parameter values $\upsilon=0.001$, $\varrho=4.6$ and $\gamma=0.225$ which correspond to the chaotic spiking region, as reported in  \cite{shilnikov2003origin}.

\section{Synchronization and chimera states}

\begin{figure}[h]
\centering
  \includegraphics[width=0.31\textwidth]{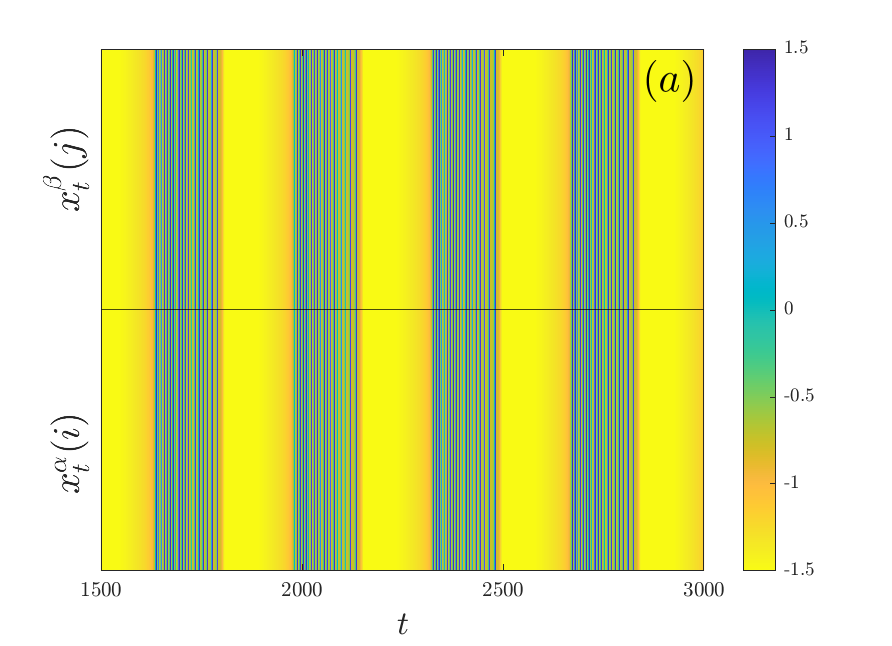} %
  \includegraphics[width=0.32\textwidth]{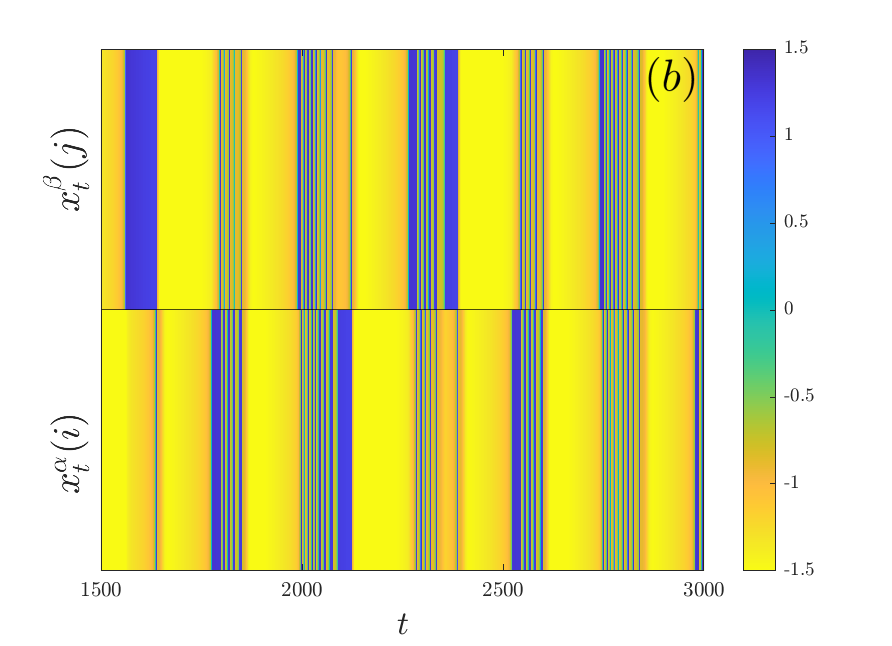} \\
  \includegraphics[width=0.31\textwidth]{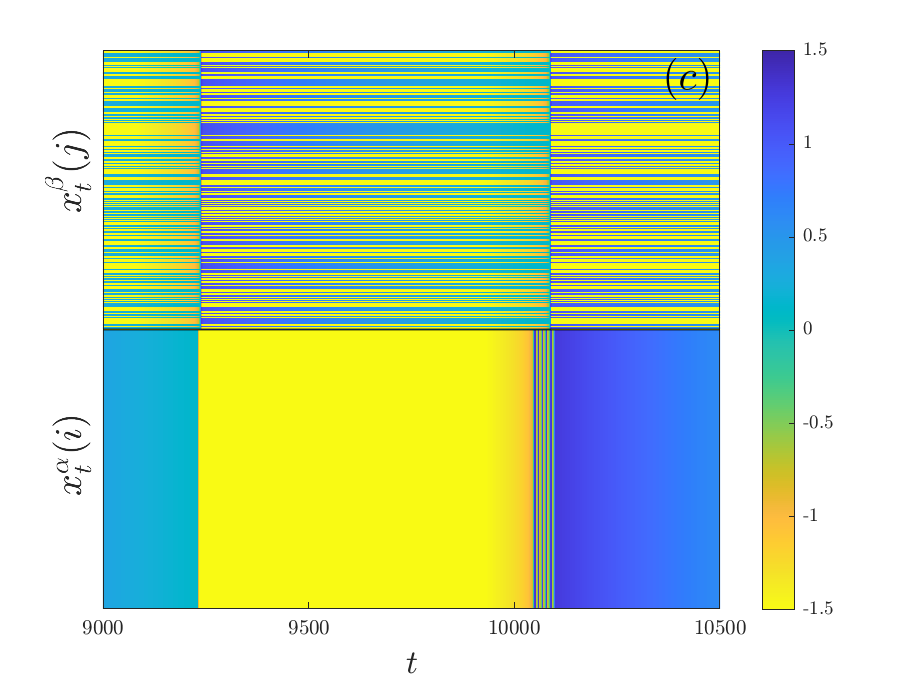} 
  \includegraphics[width=0.31\textwidth]{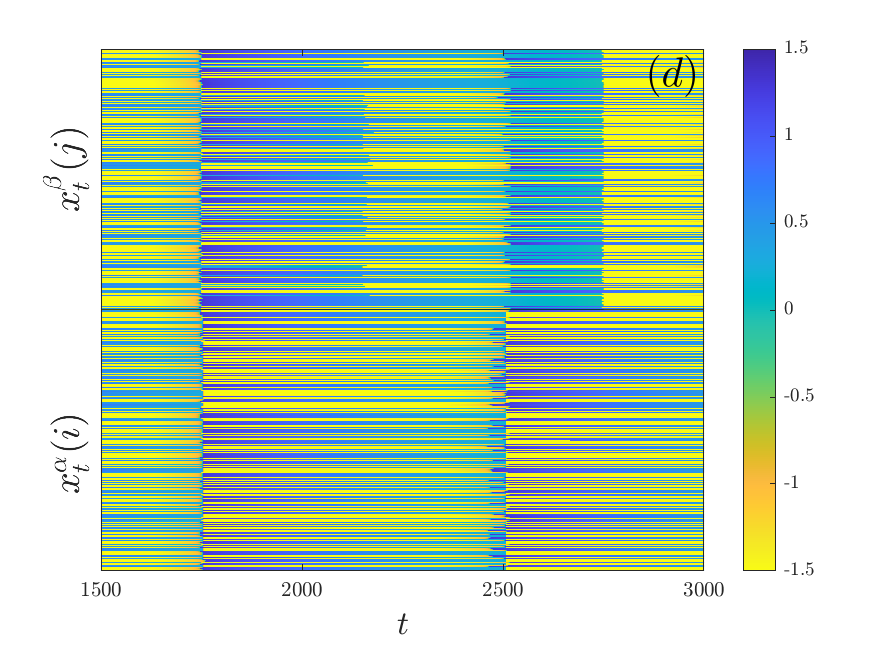}
  \caption{Spatiotemporal patterns for the two-population model, Eqs.~(\ref{ga1})-(\ref{gb2}), with local Rulkov dynamics, Eqs.~(\ref{xrul})-(\ref{yrul}). Population sizes $N^{\alpha} = N^{\beta} = 400$. Fixed parameters $\upsilon=0.001$, $\varrho=4.6$, $\gamma=0.225$. (a) Synchronization ($\langle \sigma^\alpha \rangle = \langle \sigma^\beta \rangle = 0$, $\langle \delta \rangle =0$)), $\mu=0.08$ and $\epsilon=0.04$. (b) Generalized synchronization ($\langle \sigma^{\alpha} \rangle =  \langle \sigma^{\beta} \rangle = 0$, $\langle \delta \rangle \neq 0$), $\mu=0.061$ and $\epsilon=0.02$. (c) Chimera state ($\langle \sigma^\alpha \rangle \neq 0$ and $\langle \sigma^\beta \rangle = 0$), $\mu=0.085$ and $\epsilon=0.002$. (d) Desynchronized state ($\langle \sigma^\alpha \rangle \neq \langle \sigma^\beta \rangle \neq 0$), $\mu=0.01$ and $\epsilon=0.005$.}
\label{rulkovsptp}
\end{figure}

Figure~\ref{rulkovsptp} shows the asymptotic evolution of the variables $x^{\alpha}_t (i)$, $\forall i \in \alpha$, and $x^{\beta}_t(j)$, $\forall j \in \beta$  for the two interacting population model, Eqs.~(\ref{ga1})-(\ref{gb2}), 
with Rulkov local dynamics, Eqs.~(\ref{xrul})-(\ref{yrul}), for different values of coupling parameters $\mu$ and $\epsilon$. Spatiotemporal patterns for complete synchronization, generalized synchronization, chimera state, and desynchronization are shown.

\begin{figure}[h]
\centerline{\includegraphics[width=6cm]{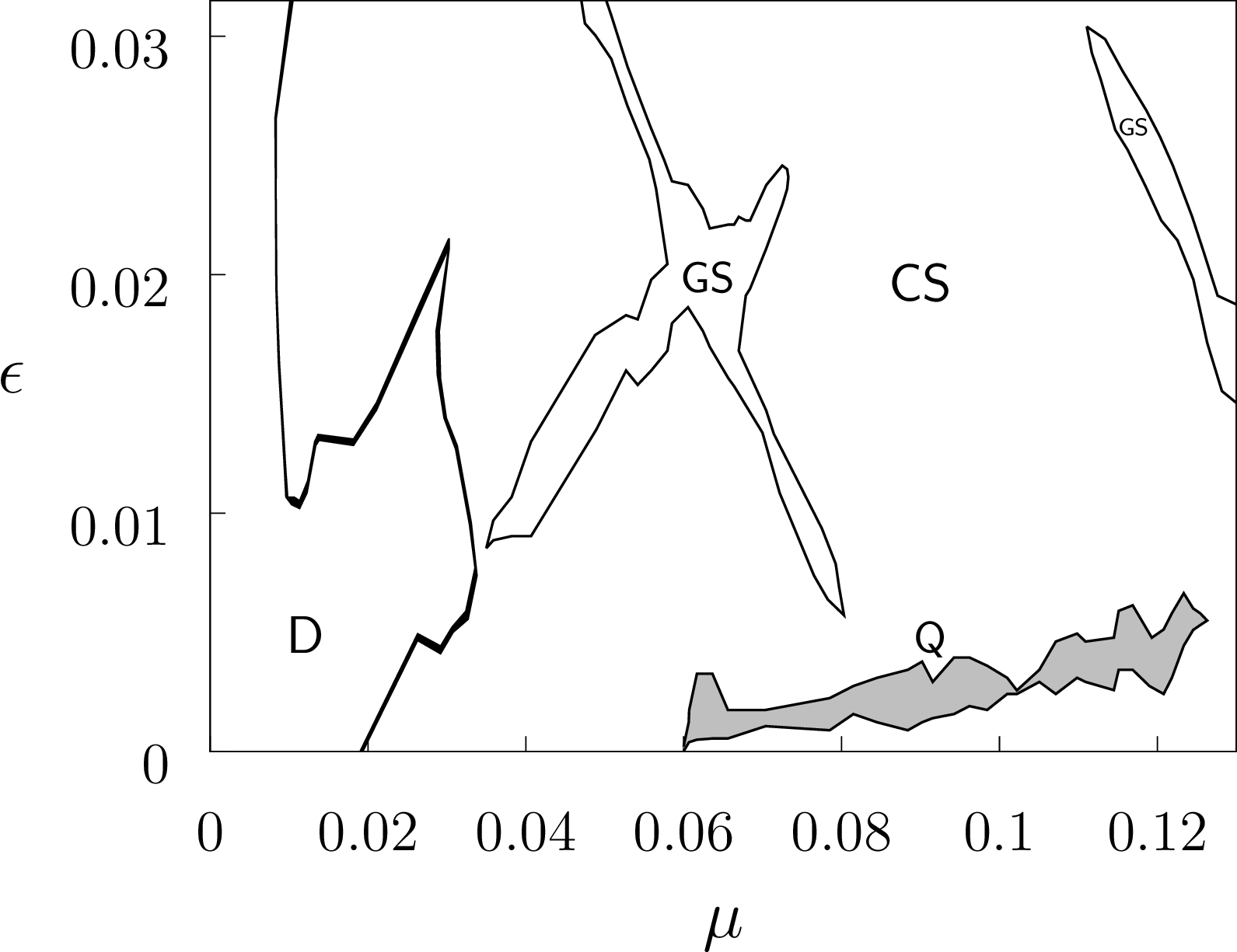}}
\caption{Phase diagram on the space of parameters $(\mu,\epsilon)$ for the two population system, Eqs.~(\ref{ga1})-(\ref{gb2}), with local Rulkov dynamics, Eqs.~(\ref{xrul})-(\ref{yrul}). Fixed local parameters $\upsilon=0.001$, $\varrho=4.6$, $\gamma=0.225$. Population sizes $N^{\alpha} = N^{\beta} = 500$. 
For each data point, the quantities $\langle \sigma^{\alpha} \rangle$, $\langle \sigma^{\beta} \rangle$, and $\langle \delta \rangle$ that characterize the different synchronization states are calculated over $1000$ iterations after discarding $3000$ transients, and averaged over $100$ realizations of random initial conditions for each population. Labels CS, GS, Q, D, indicate the regions where  collective synchronization states occur. The region where chimera states (Q) appear is colored in gray.}
\label{rulkovfig}
\end{figure}

Figure~\ref{rulkovfig} shows the collective synchronization states of the two population
system, Eqs.~(\ref{ga1})-(\ref{gb2}), with local dynamics given by the Rulkov map, Eqs.~(\ref{xrul})-(\ref{yrul}),
on the space of the coupling parameters $(\mu,\epsilon)$. A complex structure is revealed in Fig.~\ref{rulkovfig}.
Complete synchronization,
generalized synchronization, disordered, and chimera states, characterized by
the quantities  $\langle \sigma^{\alpha} \rangle$, $\langle \sigma^{\beta} \rangle$, and $\langle \delta \rangle$ appear disperse over the parameter space $(\mu, \epsilon)$.  

\begin{figure}[h]
\centering 
  \includegraphics[width=0.32\textwidth]{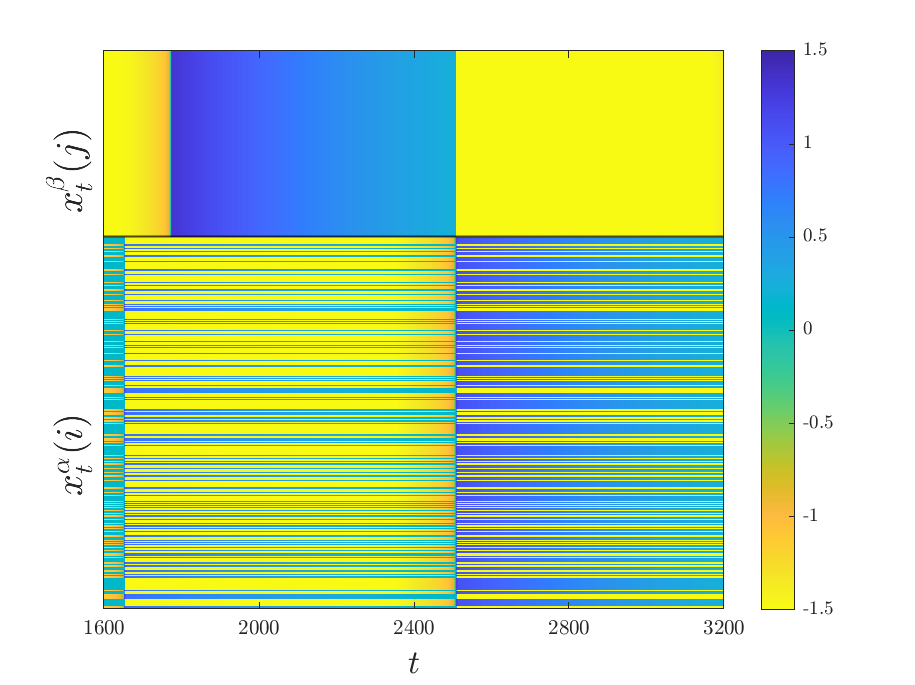}
  \caption{Spatiotemporal pattern for a chimera state in the  system, Eqs.~(\ref{ga1})-(\ref{gb2}), with different population sizes $N^{\alpha} \neq N^{\beta}$. Fixed parameters $\upsilon=0.001$, $\varrho=4.6$, $\gamma=0.225$. . 
  (a) Rulkov map for local dynamics, Eqs.~(\ref{xrul})-(\ref{yrul}), with fixed parameters $\upsilon=0.001$, $\varrho=4.6$, $\gamma=0.225$, $\mu=0.12$, $\epsilon=0.0032$, $N^\alpha=400$, $N^\beta=200$.}
\label{chimall}
\end{figure}

 To investigate the influence of asymmetry in the population sizes on the emergence of chimera states, we performed simulations for different partitions of $N^\alpha$ and $N^\beta$. Figure~\ref{chimall} shows chimera patterns for two different partitions and for the two local map-based neurons considered in this article. The relative sizes of the populations do not affect the formation of chimeras with the reciprocal global coupling scheme of the system Eqs.~(\ref{ga1})-(\ref{gb2}).

\section{Information flow between neuron populations in chimera states}
\begin{figure}[h]
\centerline{\includegraphics[width=5.5cm]{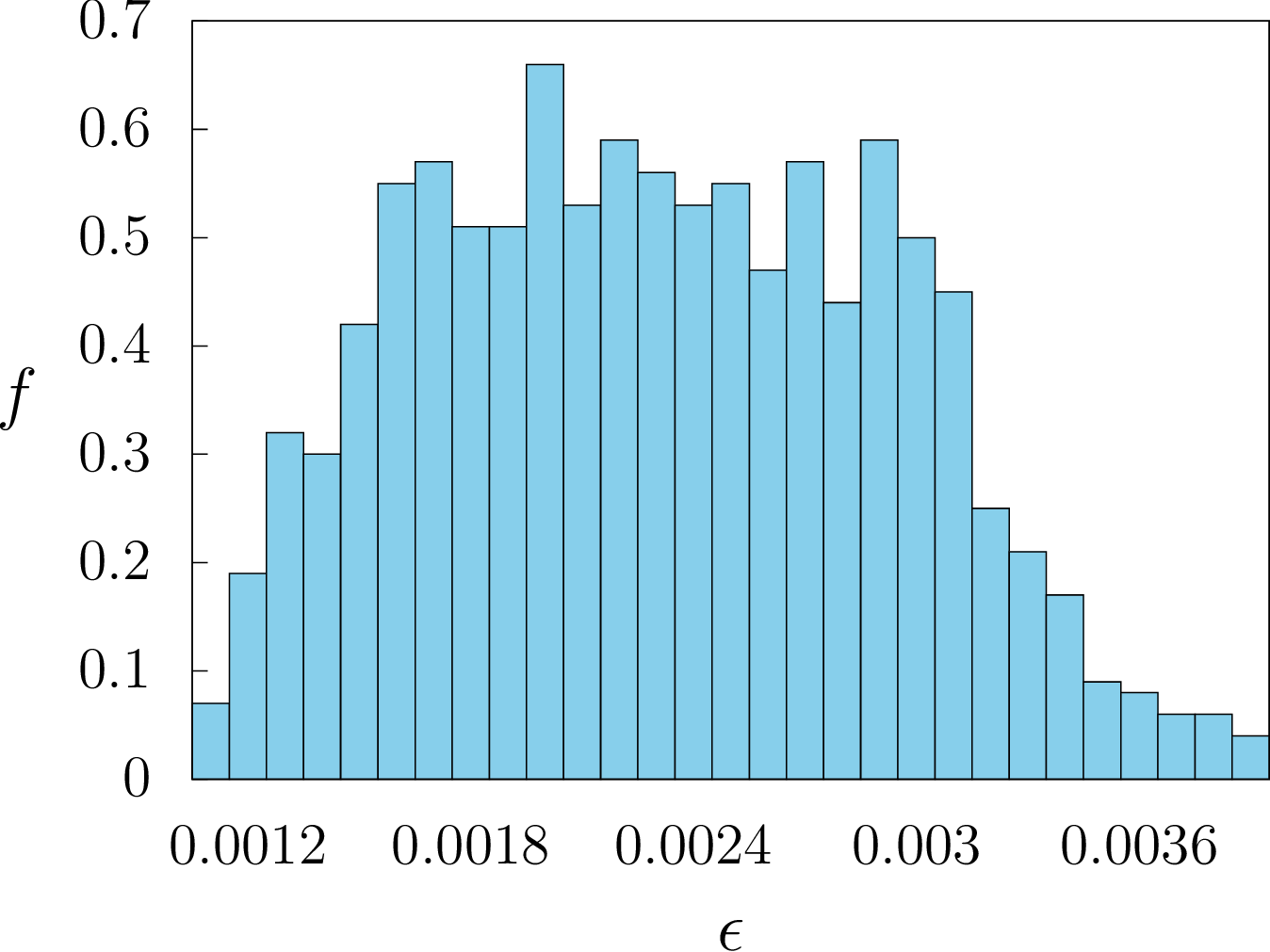}}
\caption{Frequency $f$ for the emergence of a chimera state as a function of $\epsilon$ and a fixed value of $\mu=0.09$ for the interacting populations system Eqs.~(\ref{ga1})-(\ref{gb2}).
The frequency for observing a chimera state  is calculated over $100$ realizations
of random and uniformly distributed initial conditions for each value of $\epsilon$.
 Local dynamics is given by Rulkov maps, Eqs.~(\ref{xrul})-(\ref{yrul}), with fixed parameters $\upsilon=0.001$, $\varrho=4.6$, $\gamma=0.225$. Population sizes are $N^{\alpha} = N^{\beta} = 500$.}
\label{rulkovprob}
\end{figure}

We note that chimera states are probabilistic, in the sense that their occurrence depends on initial conditions. Once formed, chimera states are stable. 
We calculate 
the frequency $f$ for occurrence of a chimera state, with one population synchronized and the other incoherent, over $100$ realizations
of random initial conditions for given parameter values.
Figure~\ref{rulkovprob} shows the frequency $f$
as a function of the inter-population coupling parameter $\epsilon$, for a fixed value of $\mu$.  We have verified that the frequency to observe a chimera state does not change as the population sizes increase.

 To elucidate the communication mechanisms between the neuron populations in a chimera state, we employ the information transfer introduced by  Schreiber \cite{Schreiber}.
 Consider two dynamical variables $y_t$ and $x_t$ interacting in a system. 
Then, the information transfer from  $y_t$ to $x_t$ is defined as \cite{Schreiber}
 
\begin{equation}
T_{y\to x}=\sum_{x_{t+1},x_{t},y_{t}} p\left ( x_{t+1},x_{t},y_{t} \right )\log\left [\frac{p\left (x_{t+1},x_{t},y_{t}\right ) p\left ( x_{t} \right ) }{p\left ( x_{t},y_{t} \right ) p\left ( x_{t+1},x_{t} \right ) } \right] ,
\label{trasn1}
\end{equation}
where $p(x_{t})$ is the probability distribution of the time series $x_t$, 
$p(x_t,y_t)$ is the
joint probability distribution of $x_t$ and $y_t$, and so on. The quantity $T_{y\to x}$
measures the degree of dependence of $x$ on $y$; i.e., the information required to represent
the value $x_{t+1}$ from the knowledge of $y_t$. Note that the information transfer is non-symmetrical
, i.e., $T_{y\to x}\neq T_{x\to y}$. 
When the two variables are synchronized, $x_t = y_t$; then 
$T_{y \to x} = 0$.
An advantage of the information transfer is that it
does not require any knowledge of the dynamical system nor prior assumptions on data generation.

In our system, Eqs.~(\ref{ga1})-(\ref{gb2}), the two populations communicate through their respective mean fields.
Thus, we consider the information transfer between the mean fields
$\bar X_t^{\alpha}$ and $\bar X_t^{\beta}$ of populations $\alpha$ and $\beta$, respectively, when a chimera state is formed. We proceed as follows. For parameter values where a chimera state appears, after a transient time of $1500$ iterations, we verify that a chimera state has formed and identify the synchronized and the desynchronized population with the labels $S$ and $D$, respectively. Then, we take a time series of 
$1500$ consecutive values for the mean fields of the synchronized and desynchronized populations, denoted by $\bar X_t^{S}$ and $\bar X_t^{D}$, respectively. From these data, we calculate the information transfers
$T_{\bar X_t^S \to \bar X_t^D}$ and $T_{\bar X_t^D \to \bar X_t^S}$ by using the definition Eq.~(\ref{trasn1}) \cite{behrendt2019rtransferentropy}.

 \begin{figure}[h]
	\centerline{\includegraphics[width=7.5cm]{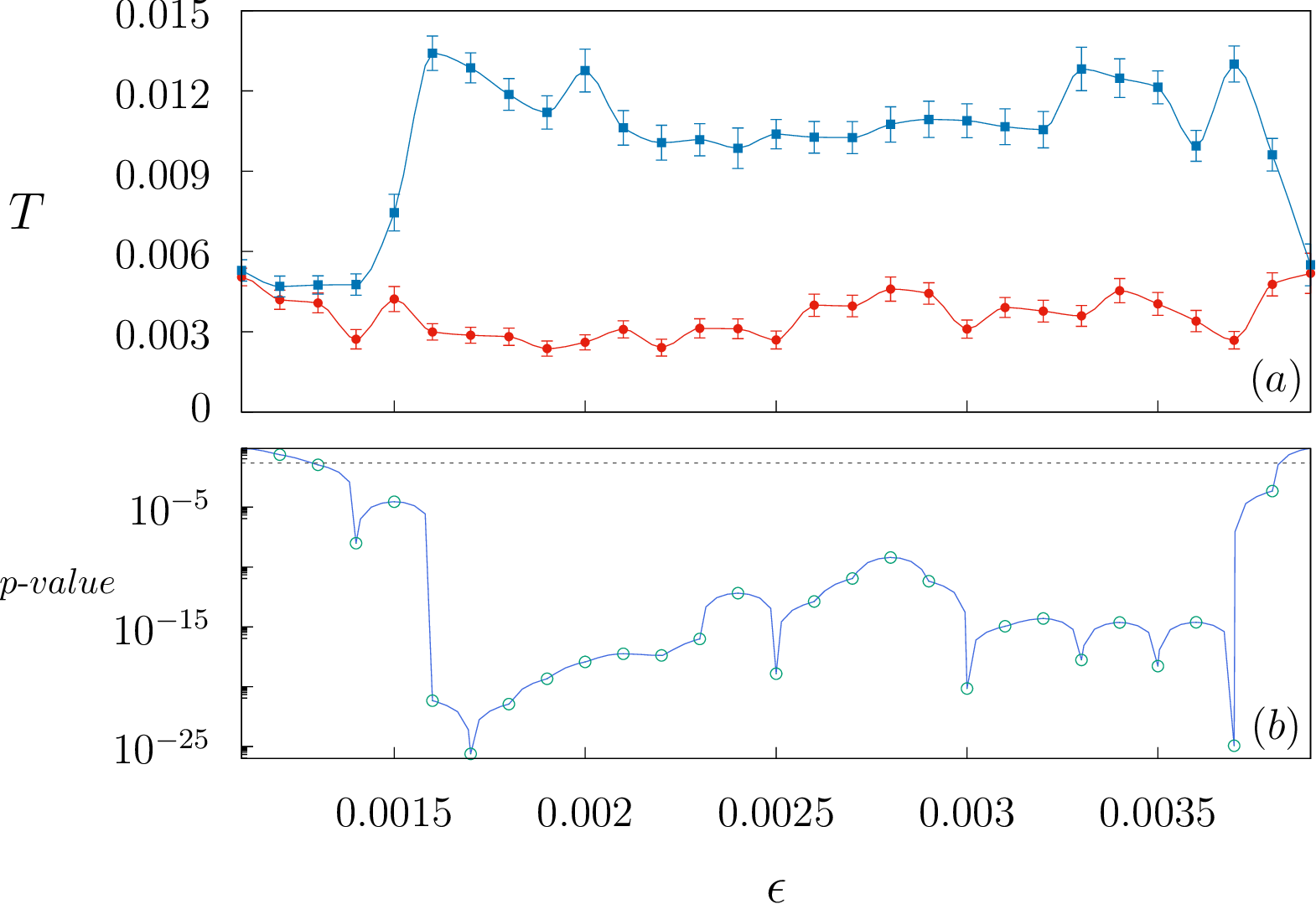}}
\caption{(a) Averaged $T_{\bar X_t^D \to \bar X_t^S}$ (squares, blue line) and $T_{\bar X_t^S\to \bar X_t^D}$ (circles, red line) in chimera states as functions of the coupling parameter $\epsilon$, for the system 
Eqs.~(\ref{ga1})-(\ref{gb2}) with fixed $\mu = 0.09$. Local dynamics is given by Rulkov maps, Eqs.~(\ref{xrul})-(\ref{yrul}), with fixed parameters $\upsilon=0.001$, $\varrho=4.6$, $\gamma=0.225$. Population sizes are $N^{\alpha} = N^{\beta} = 500$.
Each data point of $T_{\bar X_t^D \to \bar X_t^S}$ and $T_{\bar X_t^S \to \bar X_t^D}$ is the average of $100$ information transfer measures calculated over $100$ realizations of initial conditions resulting in chimera states, where the synchronized and desynchronized populations $S$ and $D$ have been identified. Error bars represent the corresponding standard errors. (b) $p$-values obtained from the Wilcoxon test as a function of $\epsilon$. The $p$-values characterize the statistical difference between  the quantities $T_{\bar X_t^D \to \bar X_t^S}$ and  $T_{\bar X_t^S \to \bar X_t^D}$. The dashed horizontal line signals the significance level $0.05$.}
\label{InfoTR}
\end{figure}

Figure~\ref{InfoTR}(a) shows the averaged quantities $T_{\bar X_t^D \to \bar X_t^S}$ and  $T_{\bar X_t^S \to \bar X_t^D}$ as functions of the coupling $\epsilon$ in the region of parameters where chimera states arise. 
Figure~\ref{InfoTR}(a) reveals that  $T_{\bar X_t^D \to \bar X_t^S} > T_{\bar X_t^S \to \bar X_t^D}$, indicating a defined direction of the flow of information between the two neuron populations in chimera states, from the desynchronized population $D$ to the synchronized one, $S$. 
We have also evaluated the statistical difference between the quantities $T_{\bar X_t^D \to \bar X_t^S}$ and  $T_{\bar X_t^S \to \bar X_t^D}$ for different values of $\epsilon$. 
To compare these quantities, we employ the Wilcoxon test  which can be applied 
when the distribution of the difference between the two means of two samples cannot be assumed to be normally distributed. 
In Figure~\ref{InfoTR}(b) we show the corresponding $p$-values as a function of $\epsilon$. The dashed horizontal line marks the significance level $0.05$.
Very low $p$-values indicate that the differences between $T_{\bar X_t^D \to \bar X_t^S}$ and  $T_{\bar X_t^S \to \bar X_t^D}$  are significant; therefore, Figure~\ref{InfoTR}(b) provides a statistical evidence that $T_{\bar X_t^D \to \bar X_t^S} > T_{\bar X_t^S \to \bar X_t^D}$
in the range of parameters where chimera states appear.

Thus, a functional directionality emerges in the system despite the structural symmetry of our model, i.e.,  population $D$ acts as the driver of population $S$ in a chimera state. 
This result is relevant in the context of epileptic seizures that have been related to chimera states in brain dynamics \cite{lainscsek2019cortical,Andrzejak2016}. It has been found
that there is a significant coupling direction in the oscillations from the thalamus to the seizure zone during
epileptic seizures \cite{li2020measuring}. 

 \begin{figure}[h]
\centerline{\includegraphics[width=8cm]{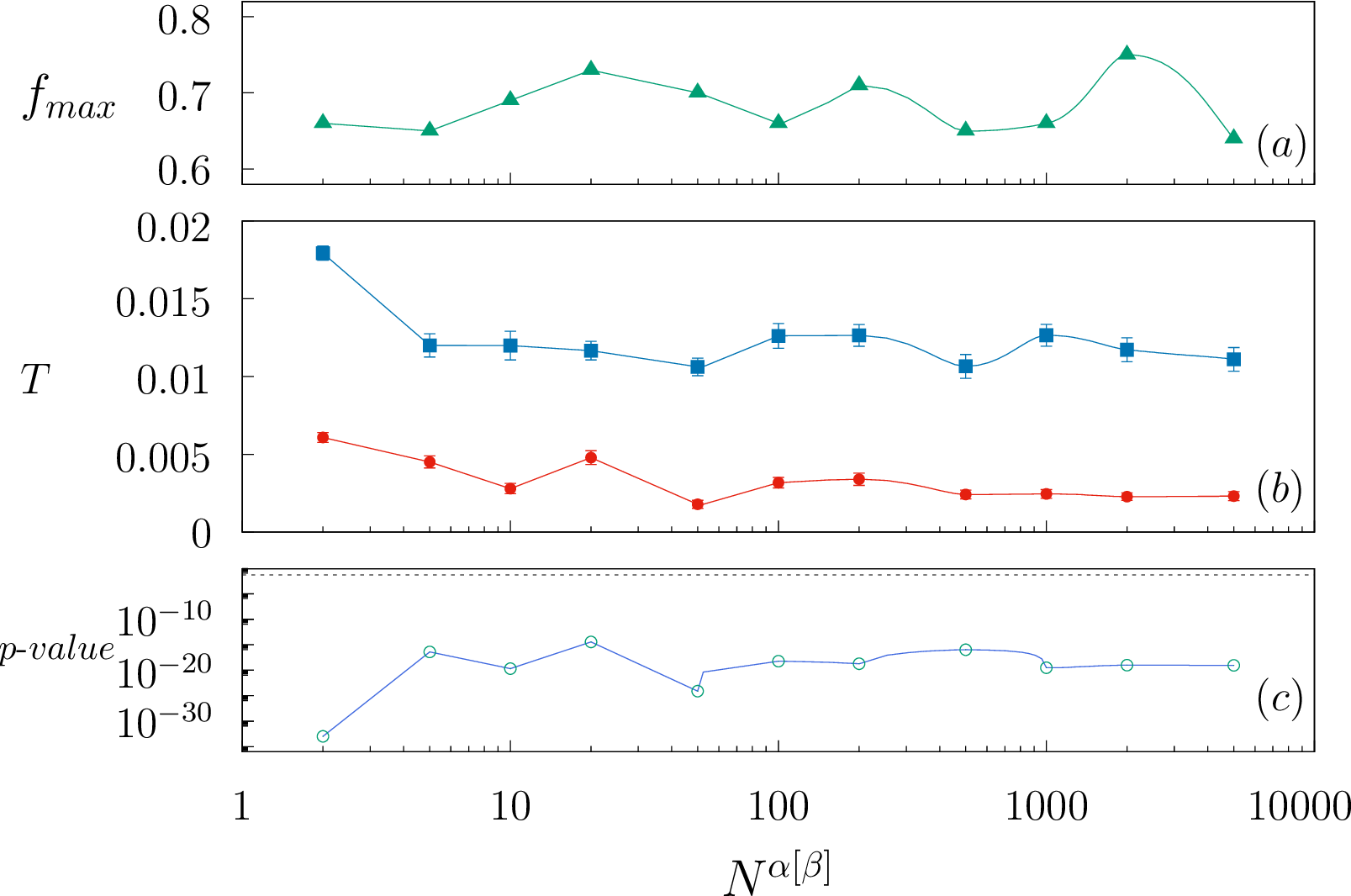}}
\caption{(a) Maximum frequency for chimera states as function of the population size $N^{\alpha}(=N^{\beta})$ in the system Eqs.~(\ref{ga1})-(\ref{gb2}).
(b) Averaged $T_{\bar X_t^D \to \bar X_t^S}$ (squares, blue line) and $T_{\bar X_t^S\to \bar X_t^D}$ (circles, red line) in chimera states  as functions of the population size $N^{\alpha}(=N^{\beta})$ in the system Eqs.~(\ref{ga1})-(\ref{gb2}).  Error bars represent the corresponding standard errors. 
(c) $p$-values obtained from the Wilcoxon test as a function of $\epsilon$. The $p$-values characterize the statistical difference between  the quantities $T_{\bar X_t^D \to \bar X_t^S}$ and  $T_{\bar X_t^S \to \bar X_t^D}$. The dashed horizontal line signals the significance level $0.05$.
Each data point on (a) and (b) corresponds to parameters $(\mu,\epsilon)$ where a chimera state has maximum frequency to occur. Local dynamics is given by Rulkov maps, Eqs.~(\ref{xrul})-(\ref{yrul}), with fixed parameters $\upsilon=0.001$, $\varrho=4.6$, $\gamma=0.225$.}
\label{probmaxsize}
\end{figure}

We have verified that the formation of chimera states subsists as the sizes of the populations are increased by several orders of magnitude.
In addition, we have investigated the influence of the size of the populations on the directionality of the flow of information in chimera states. Figure~\ref{probmaxsize}(a) shows 
the maximum frequency $f_{max}$ for chimera states as a function of the population size $N^{\alpha}(=N^{\beta})$. For each population size, we explore values $(\mu,\epsilon)$ for which chimera state arise with a frequency calculated over $100$ realizations of initial conditions. Then, we determine $f_{max}$ as the maximum value of
these frequencies. 

In Fig.~\ref{probmaxsize}(b), for each population size $N^{\alpha}(=N^{\beta})$ and its corresponding parameters $(\mu,\epsilon)$ for which the frequency of finding a chimera state is maximum, we calculate the averaged $T_{\bar X_t^D \to \bar X_t^S}$  and $T_{\bar X_t^S\to \bar X_t^D}$ by proceeding similarly as in Fig.~\ref{InfoTR}.  
The result $T_{\bar X_t^D \to \bar X_t^S} > T_{\bar X_t^S \to \bar X_t^D}$ persists independently of the system size. 
We have calculated the $p$-values 
obtained from the Wilcoxon test between  the quantities $T_{\bar X_t^D \to \bar X_t^S}$ and  $T_{\bar X_t^S \to \bar X_t^D}$  from Fig.~\ref{probmaxsize}(b).
Figure~\ref{probmaxsize}(c) shows the corresponding $p$-values as a function of $\epsilon$; they are much smaller than the significance level of $0.05$.
We note that the predominant direction of the information flow,
from the desynchronized to the synchronized population, occurs even for the smallest possible populations of neurons ($N^{\alpha}=N^{\beta}=2)$.
We have verified that the result $T_{\bar X_t^D \to \bar X_t^S} > T_{\bar X_t^S \to \bar X_t^D}$ prevails if different population partitions are employed.

\section{Conclusions}
Coupled-map network models are computationally  efficient, conceptually simple, 
and require few ingredients to study a variety of spatiotemporal processes, including synchronization phenomena, in complex systems and brain dynamics.
In this article, we have investigated a system of two interacting populations of coupled maps subject to reciprocal global interactions, where the local units in each population exhibit neuron-like dynamics. We have focused on  parameter values where the local maps exhibit chaotic spiking behavior. It has been found that chaotic dynamics in neurons leads to a greater diversity of responses that are  relevant for neural computations \cite{rasmussen2017chaotic}.

We have characterized four collective synchronization states emerging in the system: (i) complete synchronization, where the populations are synchronized to each other, (ii) generalized synchronization, where each population is synchronized, but no to the other, (iii) chimera state, where one population is synchronized and the other remains incoherent, and (iv) desynchronization, where both populations are desynchronized.

Chimera states are probabilistic in the sense that their occurrence depends on initial conditions. On the parameter regions where the chimera states arise, the dynamics of the two-population system is multistable; i.e., several attractors coexist for the same parameter values. This behavior is typical of phase transition regions in dynamical systems. Chimeras represent an intermediate state between ordered and disordered phases.

We have found a well-defined direction of the flow of information in chimera states, from the desynchronized neuron
population to the synchronized one. This result is independent of the population sizes or population partitions. The incoherent population functions as a driver of the coherent neuron population in a chimera state.  This finding is consistent with previous reports where a measure of transfer entropy, applied to epileptic electrocorticography data,
shows a significant coupling direction from the anterior nucleus of thalamus to the synchronized seizure onset zone during seizures \cite{li2020measuring}. 

Although our results are obtained for specific neuron-map local dynamics, we expect that the  drive-response causal relationship between the desynchronized and synchonized subsets  persists 
for chimera states arising in general dynamical networks. 

Models such as the presented here, where specific coupling parameters give rise to synchronization and chimera states, can provide insight to study 
phenomena such as unihemispheric sleep and  brain disorders associated to alterations in the neural synchrony activity within and across brain areas \cite{wang2020brief}. 
Neurological disorders, such as epilepsy, Alzheimer's and Parkinson's disease,  and autism, have been related to abnormal neural synchronization patterns \cite{uhlhaas2006neural, delbeuck2003alzheimer, babiloni2017abnormalities,jiruska2013synchronization,dinstein2011disrupted}. 

Map-based neuron models are gaining recognition in the field of neuroscience, where they have been implemented in several studies \cite{nowotny2005self,yu2016functional,sayari2023analyzing}.
Although the two population model considered here is not specific in terms of biological plausibility, it yields the emergence of rich collective behavior that can be a useful tool in computational neuroscience.
Future extensions of the two interacting population model proposed here should include the investigation of time delays, heterogeneity in the local dynamics, different connectivity networks, the similarity of collective states in driven and in autonomous systems, and the influence of external inputs.

\section*{Acknowledgments}
V.J.M.R. thanks Doctorado en Ciencias de la Complejidad Social, Universidad del Desarrollo, Chile, for a Ph.D. scholarship and support. M.G.C. thanks ViceRectorado de Investigaci\'on e Innovaci\'on, Universidad Yachay Tech, Ecuador, for support.

\bibliography{chimerabiblio}

\end{document}